# Endogenous and microbial volatile organic compounds in cutaneous health and disease

Emer Duffy, Aoife Morrin*

Insight Centre for Data Analytics, National Centre for Sensor Research, School of Chemical Sciences, Dublin City University, Glasnevin, Dublin 9, Ireland

*Corresponding author; aoife.morrin@dcu.ie

**Abstract**

Human skin is a region of high metabolic activity where a rich variety of biomarkers are secreted from the stratum corneum. The skin is a constant source of volatile organic compounds (VOCs) derived from skin glands and resident microbiota. Skin VOCs contain the footprints of cellular activities and thus offer unique insights into the intricate processes of cutaneous physiology. This review examines the growing body of research on skin VOC markers as they relate to skin physiology, whereby variations in skin-intrinsic and microbial metabolic processes give rise to unique volatile profiles. Emerging evidence for volatile biomarkers linked to skin perturbations and skin cancer are examined. Microbial-derived VOCs are also investigated as prospective diagnostic markers, and their potential to shape the composition of the local skin microbiota, and consequently cutaneous health, is considered. Finally, a brief outlook on emerging analytical challenges and opportunities for skin VOC-based research and diagnostics is presented.

1. Introduction

The skin is the first line of defense between the human body and its external environment, preventing loss of moisture and entry of pathogenic organisms. It is a region of high metabolic activity where secretion of a rich variety of biochemical markers occurs throughout the stratum corneum (SC), including lipids, peptides, proteins, cytokines, nucleic acids and volatile organic compounds (VOCs). It is also an ecosystem, home to diverse microbial communities that influence human health and disease, and which secrete their own metabolites including toxins, antibiotics and VOCs. Local skin biochemistry offers a valuable insight into health as it can reflect various pathologies including localised diseases like atopic dermatitis, psoriasis and cancer [1], and even systemic conditions such as diabetes and cardiovascular disease [2].

Skin volatile emissions are attracting considerable scientific and clinical interest in recent years as they contain the footprints of cellular activities and can thus reflect the metabolic condition of an individual, offering a non-invasive route to probe the body's biochemistry [3]. Skin VOCs are derived from apocrine, eccrine and sebaceous gland secretions, and their interactions with microorganisms on the surface. Different regions of skin emit specific VOCs generating a complex mixture of compounds often referred to as human odour [4] comprising a diverse group of carbon-based molecules that are volatile at ambient temperatures. A compendium of VOCs from the human body was recently published, outlining a total of 1840 compounds from skin, breath, blood, urine, feces, saliva and milk. Skin secretions were assigned 532 compounds [5] and featured a variety of compound classes including hydrocarbons, carboxylic acids, aldehydes, alcohols, esters, ketones and amines. Skin volatiles are generating significant interest in numerous fields, from cosmetics (*e.g.* in the design of fragranced products such as perfumes and deodorants) [6],[7], to diagnostics [8], forensics [9], safety and security applications (*e.g.* search and rescue) [10], monitoring psychological stress [11], and ecology of blood-sucking insect vectors of human disease [12]. Recent research has highlighted the link between skin volatiles and the potential passage of compounds from blood vessels [13], dietary influences [13],[15] and age-related metabolic activity [16]. Environmental factors and personal habits can also influence skin volatile composition, including level of physical activity, personal hygiene, and cosmetic or fragrance usage [9]. Monitoring skin volatiles offers a non-invasive route towards probing pathophysiological status and monitoring therapy, and research in this area to date has primarily focused on internal or systemic disease states for which consistent data has emerged on clinically important skin volatiles, including diabetes [17], hepatic disease [18] and malaria [19].

Skin volatile profiles reflect metabolites of the dermal and epidermal cellular layers and the compounds carried to and from the skin *via* the blood stream. They also reflect the volatile metabolites derived from the skin flora and the environmental volatiles on the surface of skin. In cases of perturbation to the skin (*e.g.* barrier disruption, tissue damage or cancer) the volatile profiles may be changed by alterations in cellular metabolism. Volatile profiles may also be altered by the loss of normal skin cellular layers and their metabolites, or by metabolites associated with the processes of infection and healing. Consequently, skin VOCs can provide a unique insight into the intricate processes of cutaneous health and disease. There is a growing body of research highlighting the value of this approach to reveal important biochemical information about the skin, wherein variations in skin intrinsic metabolic processes and commensal microbial populations generate unique volatile profiles. Characteristic volatile profiles have been reported for chronic wounds [20], melanoma [21], compressed tissue [22], and skin barrier impairment [23] thus volatile analysis can reveal important information about the health of the skin itself. This approach could be invaluable for monitoring barrier function, tissue damage or skin/wound infection and healing, where a non-invasive approach could reduce reliance on surgical excisions. Furthermore, it could enrich understanding of underlying skin metabolism and physiology in a non-destructive manner with applicability to cell culture and human research alike. This review will provide an overview of emerging research on volatiles as they relate to skin physiology, where the rich biochemistry of skin volatile secretions can be exploited to develop non-invasive approaches to measure skin properties, monitor skin disorders, evaluate efficacy of treatment, or detect disease or infection at an early stage.

## 2. Discussion

*2.1 Sampling and analysis of skin volatile emissions*

Studying skin VOCs requires a high degree of analytical sensitivity due to the low concentrations (low to sub-ppb) emanating from skin. This has led to a reliance on mass spectrometric methods for their analysis. A detailed overview of analytical techniques for VOCs from the human body is provided in a recent review [24]. Briefly, techniques for their analysis include gas chromatography coupled with mass spectrometry (GC-MS) and direct MS methods such as selected ion flow tube mass spectrometry (SIFT-MS) [13], proton transfer reaction mass spectrometry (PTR-MS) [25], membrane inlet mass spectrometry (MIMS) [26] and secondary electrospray ionization mass spectrometry (SESI-MS) [27]. Gas chromatography has been coupled with ion mobility spectrometry (IMS) for near real-time characterization of skin volatiles [28],[29], however, parallel GC-MS analysis is vital to reliable compound identification as IMS does not have a commercially available substance library. Electronic noses (e-nose) have also been employed for skin VOC detection and recognition in armpit odours [30] and melanoma [21]. An e-nose comprises an array of chemical sensors (*e.g.* electrochemical, gravimetric or optical sensors) for detection and classification of VOCs. It offers a miniaturized, sensitive and rapid approach to the analysis of volatile mixtures without the need for their separation into individual components but provides limited identification. E-noses are prone to signal drift and reduced sensitivity in the presence of water vapour or high concentrations of a single analyte [31].

GC-MS is the gold standard technique for investigation of trace level VOCs, with advanced identification mechanisms and a capacity to analyse hundreds of species simultaneously. Headspace solid-phase microextraction (SPME) with GC-MS has emerged as the method of choice in skin volatile research, owing to the sensitivity and identification capabilities of MS coupled with the ease of performing headspace (HS) sampling with SPME [16],[23],[32],[33],[34]. The determination of skin volatiles often requires a pre-concentration step to enrich the target compounds to fit within the range of detection of the analytical system. HS-SPME pre-concentration is sensitive, non-invasive, solvent free and easily performed *in vivo,* reducing reliance on more labour intensive methods like bag sampling. It enables trapping of volatiles on adsorbent coated fibres followed by direct thermal desorption in a GC injector, and can be used for direct sampling of skin VOCs, or to extract a sample collected through an intermediary (*e.g.* cotton or gauze pads) in contact with the skin. A review of techniques for sampling skin VOCs is provided by Dormont *et al.* [35]. Briefly, these methods include solvent extraction [16] and VOC collection onto adsorbents in headspace or contact mode (*e.g.* glass beads, cotton gauze, polydimethylsiloxane membranes) [12],[36],[37]. More recently, polydimethylsiloxane tubing has been employed as a wearable passive sampler looped around the wrist or ankle [38]. Volatiles can be recovered from these phases for subsequent analysis *via* thermal desorption, solvent extraction, or collection onto an adsorbent trap (*e.g.* porous polymer adsorbent) using an airflow process. Headspace sampling is preferable, since contact sampling can introduce interferents (*e.g.* dust, lipids, non-volatiles) to the sample, which can contaminate the GC injector and column [37]. Numerous studies have included some degree of control over skin pre-treatment prior to sample collection to reduce environmental contributions. Participants are generally requested to avoid using cosmetics on the day of sampling, and to wash the skin site with water, or with fragrance-free soap prior to sampling [9],[39]. However, the influence of cosmetics is generally still an issue, with some compounds persisting in excess of 2 weeks in the skin [12]. Other studies have included a degree of control over dietary factors [40] or introduced an exercise component to generate sweat prior to sampling [33].

*2.2 Endogenous skin volatile emissions in vivo and in vitro*

Over 500 volatiles have been reported isolated from skin secretions [5]. Studies employing headspace sampling of skin typically recover between 20-90 VOCs that are airborne at normal body temperature [35] with recovered compounds spanning a variety of chemical classes including hydrocarbons, aldehydes, alcohols, carboxylic acids, esters, ketones and amines. Skin volatiles predominantly derive from gland secretions and the metabolism of the skin microbiota. The skin glands are located in the dermis and terminate in secretory ducts that open on the skin's surface. Gland distribution at the skin surface varies across the body and is partially reflected in the regional diversity of skin volatiles. Apocrine glands secrete sweat; a turbid fluid containing water, lipids, proteins, and odour precursors [41]. They are localized in a few regions of skin and contribute greatly to VOC emissions produced from the axillary regions which also harbor a diversity of microbial flora. Axillary volatiles primarily contain $C_{6-11}$ carboxylic acids and alkanes [35]. The sebaceous glands secrete sebum which has the function of lubricating the skin. Sebum is composed of triglycerides, wax esters, squalene and free fatty acids. Comparison of volatiles in sites of differing sebaceous gland density revealed a considerable number of overlapping compounds, and some notable differences relating to the site of emanation [16]. Eccrine glands are distributed across the skin and have their functions in excretion and thermoregulation by secreting water and electrolytes. They also play a role in maintaining skin homeostasis by secreting moisturizing factors such as lactate, urea, sodium and potassium to maintain skin hydration [42]. Eccrine secretions originate in the extracellular interstitial fluid, and therefore reflect blood plasma chemistry

[43]. They are distributed across the body and contribute to VOC mixtures from numerous regions on the skin.

Some of the most frequently reported volatile compounds recovered from skin include nonanal, decanal, 6-methyl-5-hepten-2-one and 6,10-dimethyl-5,9-undecadien-2-one (geranyl acetone), followed by hexanal, octanal, benzyl alcohol, (*E*)-2-nonenal, undecanal and hexadecane [35]. Volatile profiles from feet are often dominated by short chain fatty acids (SCFAs) whereas hand and forearm volatiles tend to exhibit a greater distribution of aldehydes and ketones [16],[44],[45],[46]. Several VOCs have been proposed as markers of increasing age including (*E*)-2-nonenal [47], nonanal, dimethylsulfone and benzothiazole [16]. Numerous gender marker compounds have also been proposed, (including pentadecanoic acid, hexadecenoic acid, heptadecanoic acid, nonadecane and docosane) which permitted discrimination between the sexes in a study on axillary emanations collected from 89 men and 108 women using PDMS-coated stir bars [48]. Variations between individuals' volatiles due to microbial influences have also been reported [35],[44].

A library of endogenous VOCs has been proposed based on volatile profiling of human cell lines [49]. Analysis of VOCs in human primary cells, cell lines and cultures of microorganisms is expected to provide invaluable insights into the biochemistry of volatile compounds observed in humans. Complementary studies on skin VOCs from different sources (tissue and culture studies) will assist in validating the origin of candidate compounds and their potential association with health or disease. HS-SPME sampling offers a non-invasive method of biochemical interrogation that is repeatable and easily performed on cell culture, tissue and human participants. Our group have recently demonstrated an HS-SPME method for characterization of volatiles *in vitro* and *in vivo* [50]. A total of 16 compounds were reliably identified from HS-SPME GC-MS analysis of 3D human skin equivalents (HSEs). A similar variety of compound classes to that present in normal skin was observed (with the addition of terpenic compounds camphene, camphor and isoborneol), however the individual compounds and their relative distribution varied substantially between HSEs and human skin. Esters comprised the largest portion of the HSE volatile profile, followed by acids, aldehydes and terpenes. There were 3 common compounds in HSEs and participants (octanal, nonanal, and 2-ethyl-1-hexanol) and several other VOCs present in HSE samples which were reported elsewhere in human participants (nonanol, camphor and n-hexadecanoic acid). A number of compounds identified were attributed to exogenous sources, including ethylbenzene, styrene and 2-ethyl-1-hexanol. Volatile profiles in monolayer (2D) and 3D matrix immobilized cultures of human dermal fibroblasts were compared using HS-SPME GC-MS by Acevedo *et al*. 6 compounds (cyclohexanol, styrene, benzaldehyde, ethylhexanol, acetophenone and 1,3-di-*tert*-butylbenzene) were identified, however, all 6 compounds were also present in the medium control samples and therefore unlikely to be endogenous in origin [51]. Numerous compounds identified in cellular VOC studies appear not to be from cellular metabolism but are instead derived from exogenous sources such as the growth medium or environmental contaminants [52]. The detection of metabolic changes *via* skin VOCs will be enabled by a combined knowledge of the cellular source and the fundamental biochemical process, which demands careful characterization of VOCs *in vitro* and *in vivo*.

2.2.1 Skin perturbations alter volatile emissions

Skin volatiles reflect metabolites of the dermal and epidermal cellular layers and the compounds carried to and from the skin *via* the blood stream. In cases of skin perturbation, the volatile profiles may be changed by alterations in cellular metabolism. Prolonged compression can limit the availability of oxygen in a region of tissue under mechanical load [53]. Early detection of compression tissue damage *via* volatile markers from the skin was proposed by Dini *et al.* [22]. They employed a cotton patch sampling approach with subsequent extraction by SPME and analysis by GC-MS and electronic nose. Compressed and non-compressed regions of skin were sampled in 9 healthy participants and 9 hospitalized intensive care patients. Differences were observed in the chromatographic pattern of volatiles for compressed and non-compressed tissue (Table 1). However, differing volatile emissions were observed for hospitalized patients and healthy participants, irrespective of the condition of the tissue. This was attributed to differences in the medication and environment of the 2 groups. Further research is needed to understand these environmental effects (hospital *vs* other settings) and to elucidate volatile markers representing the early onset of pressure ulcers. This could inform development of a portable diagnostic approach, such as an electronic nose, with sensing materials specific to those markers permitting early detection of tissue damage prior to pressure ulcer occurrence.

Skin perturbation *via* acute disruption to the skin barrier induces obvious epidermal hyperplasia and inflammation, and can result in accelerated lipid synthesis, increased epidermal DNA synthesis and cytokine production [54]. Our research group have demonstrated that skin barrier disruption can impact the volatile emissions from skin [23]. The volatile profile of healthy skin was investigated before and after acute barrier disruption through tape stripping of the stratum corneum. Tape stripping is a simple, robust method used in skin physiology research and as an in vivo model for skin diseases characterized by skin barrier weakening, such as atopic dermatitis. Discriminating volatile profiles were observed for all 7 participants before and after acute barrier disruption. Volatile emissions revealed the protective hydro-lipid film that functions within the skin barrier was impacted, with compounds from sebaceous components and their oxidation products (e.g. squalene, squalene, octanal, nonanal) showing substantial changes after barrier disruption (Table 1). Glycine up-regulation was also observed and may indicate an impact on the skin's natural moisturizing factor. This approach could enrich understanding of skin barrier function and treatment efficacy, which are key to improving outcomes for chronic conditions like atopic dermatitis and psoriasis. The observed up-regulation of exogenous compounds (linalool and phenethyl alcohol) in the study underlines the persistence of exogenous compounds on the skin and suggests the skin volatilome might provide a useful source of information regarding an individual's environment and exposures. Volatile screening offers the potential to detect changes in the skin that are undetectable in the traditional palpable and visual assessment, it could reduce reliance on surgical excisions and may provide opportunities for early detection of tissue damage, barrier impairment, or irritant reactions for application in areas such as clinical diagnostics, physiology research and cosmetic and skincare testing.

2.2.2 Emerging volatile markers for skin cancer

There have been a number of anecdotal reports of canine olfactory detection of skin cancers (melanoma and more recently basal cell carcinoma) which has motivated the training of dogs as a diagnostic tool [55], [56],[57]. To date this research has primarily focused on melanoma, which is represented in a low percentage of skin cancer cases (~5%) but is classified as the deadliest form of skin cancer, accounting for up to 75% of all skin cancer deaths [58]. Dogs trained to sniff patients were correct in identifying 75-86% of melanomas hidden beneath bandages [59]. This suggests the emission of unique volatile metabolic products of melanoma pathophysiology that differ from that of healthy skin or moles. Cancer cells have altered metabolisms, which are expected to produce different profiles of metabolites and this has motivated a growing interest in developing a non-invasive diagnostic approach around skin volatiles [21],[52],[60]. Currently the gold standard diagnostic approach for melanoma involves excisional biopsy and histological examination of suspicious lesions. There is a demand for more objective and less-invasive examination methods to support clinicians in determining when to perform a biopsy [61]. Volatile screening could permit early detection of melanoma in a non-invasive manner which may increase the chance of successful treatment.

Discovery of differences in volatiles between melanoma and normal skin is best accomplished using GC-MS, however, translation of such an approach into a clinical setting requires user-friendly portable analytical technology (*e.g.* e-nose). Studies employing gas sensor arrays (comprising 7 quartz crystal microbalance (QCM) chemical sensors, each coated with a different metalloporphyrin) and GC-MS have reported that melanoma-related volatiles differed from those of normal skin, however, compound identities were not reported for volatiles that could discriminate between melanoma and normal skin [21]. PLS-DA modelling of sensor array data collected from 40 participants (including 9 melanoma) revealed a degree of separation between naevi and melanoma, with correct prediction of melanoma reported for 70% of cases. GC-MS characterization of volatile emissions in 2 of these participants (comparing naevi, normal skin and melanoma) revealed subtle differences in the chromatogram relating to skin cancer, with 1 compound identity suggested (propanal). Any other relevant compounds could not be elucidated due to low relative abundances [21].

More recent research is shedding light on the composition of volatile emissions from melanoma, naevi and normal skin. HS-SPME GC-MS characterization of volatiles in punch biopsy melanoma and normal skin samples revealed differences in the volatile signatures, with up-regulation of lauric acid and palmitic acid reported for melanoma samples [62]. However, numerous other compounds reported from GC-MS analyses of melanoma volatiles do not appear to be from cell metabolism. Compounds such as methoxy-phenyl oxime, butylated hydroxytoluene [63], isopropyl palmitate, propylene glycol, 2-ethyl-1-hexanol and styrene [60] are all likely derived from exogenous sources. Control sampling and examination of corresponding mass spectra should permit elimination of compounds arising from environmental or instrumental artifacts (including column bleed SPME fibre volatiles, background air and cosmetics). After eliminating such contaminants from their chromatographic data, Kwak *et al.* discovered differences in volatile sulfur compounds in the headspace of cultured melanoma cells and normal melanocyte cells using SPME with GC-MS. They investigated volatile emissions from three types of primary melanoma cells (radial growth phase, vertical growth phase and metastatic melanoma cells) as well as normal neonatal foreskin melanocytes. Melanoma cells grown in culture produced dimethyl disulfide and dimethyl trisulfide, two volatiles which were not present in those emitted by normal melanocyte cultures [52]. Furthermore, dimethylsulfone was seen in greater amounts in metastatic melanoma cells compared to normal cells, and metabolic differences between these two cell types in the processing of benzaldehyde was proposed. Detection of melanoma biomarker dimethyl disulfide using cavity ringdown spectroscopy at 266 nm has been investigated as an alternative diagnostic approach, with potential for sub-ppb detection limits reported [64]. Application of a gas sensor array comprising single walled carbon nanotubes functionalized with single stranded DNA to cell culture supernatants demonstrated distinguishing differences in volatiles from normal and melanoma cells, and from different types of melanoma cells. [52] Both canine olfactory recognition of melanoma and emerging evidence of differences in volatiles from melanoma and normal skin (cells and biopsies) offer promise to advancing the early detection of melanoma using non-invasive analytical techniques. Careful elucidation of endogenous volatile biomarkers with consideration of environmental contaminants will inform development of selective detection strategies suitable for translation to clinical applications.

*2.3 Cutaneous microflora and microbial volatile emissions*

Human skin is colonized by a large diversity of microorganisms collectively known as the skin microbiota. The skin microbiota is made up of complex and dynamic communities of microbes including bacteria, viruses and fungi. Microbial profiling *via* 16S rRNA sequencing has revealed the presence of highly diverse communities across distinct topographical skin sties. These communities are generally dominated by bacteria belonging to 3 main phyla (Actinobacteria, Bacteriodes and Firmicutes) [65]. There is growing recognition around the cutaneous microflora as an important component of skin health, as well as many disease processes [66]. Microbial etiological agents are associated with various skin disorders, including *Staphylococcus aureus, Staphylococcus epidermidis* and *Propionibacterium acnes* [67],[68] and dysbiosis in the skin microbiota has been associated with cutaneous inflammatory disorders including psoriasis, atopic dermatitis (AD) and rosacea [65],[69],[70]. However, the majority of bacteria resident on the skin are non-pathogenic commensals that can be beneficial for their host in numerous ways, such as through immune system regulation [71] or elimination of competing pathogens through production of toxic compounds like bacteriocins or antibiotics [69],[72]. As part of their normal metabolism, microorganisms produce VOCs, and microbial volatile organic compounds (mVOCs) comprise a wide variety of compound types including fatty acids and their derivatives (*e.g.* hydrocarbons, alcohols, ketones), nitrogen- and sulfur-containing compounds, terpenoids and aromatic compounds [73]. The composition of volatiles produced by a microbial community is dependent on the diversity and types of species present and the environmental conditions including the availability of carbon energy sources, and conditions like pH and temperature [74]. HS-SPME GC-MS screening of VOC profiles from human skin microbiota representatives growing under diverse conditions *in vitro* revealed differential production of volatiles depending on bacterial strain and growth conditions (incubation temperature and media composition) [75]. The majority of volatiles observed from bacterial cultures under investigation were alkenes, alcohols, ketones and aldehydes (products of fatty acid biosynthesis). Volatiles produced by *Pseudomonas aeruginosa* were dominated by 1-undecene and an increase in the prevalence of long chain methyl ketones (*e.g.* 2-undecanone) was observed when incubation temperature was raised from 30 to 37 °C. The 3 strains of *Staphylococcus epidermidis* investigated all produced significant amounts of long chain methyl ketones with fewer alcohols, aldehydes and alkenes present. However, differences in the breakdown of VOCs between strains were apparent, with 3-methylbutanal production observed in only 1 of 3 strains. Among the other bacterial strains investigated, *R erythopolis* produced no identifiable VOCs, and low levels of VOCs were recovered from *B. epidermidis* and *C. xerosis* including 2-butanone and 2-acetylthiazle [75].

In addition to producing VOCs as part of their normal metabolism, the skin microbiota also produces VOCs by metabolizing secretions from our skin glands, which can have a major impact on the composition of our VOC profile (*i.e.* our body odour). The topographical diversity in the microbiota and the distribution of glands in different skin sites results in variations in the volatile profile with emission site [52]. *Corynebacterium* species degrade precursor compounds found in odourless apocrine secretions into volatile fatty acids and thioalcohols that are associated with axillary malodour [76], while *Staphylococcus epidermidis* degrades the leucine found in sweat to produce isovaleric acid, a major component of foot odour [77].

There is a rapidly growing interest in the investigation of VOC profiles from different skin commensal bacterial species [75],[78]. This research is being driven by a number of motivations including understanding the bacterial transformation of skin secretions into odiferous VOCs [76],[79],[80] and uncovering the role of bacterial VOCs in interactions with other organisms [81],[82],[83] as well as bacteria-bacteria interactions [78]. Moreover, there is a focus on the potential use of VOCs from pathogenic bacteria as diagnostic tools for investigating microbial infections [84]. The detection of microbial VOCs (mVOCs) has long been used as a means of identifying bacterial species [85] and it is anticipated that this approach could permit non-invasive investigation of skin microbial communities *in situ* [84]. Emerging research around mVOCs in cutaneous health and disease will be discussed herein with a focus on their potential as non-invasive diagnostic tools for microbial associated cutaneous disease.

2.3.1 Investigating volatile profiles in different skin commensal bacterial species

The local skin microbiota plays an important role in supporting the skin's function as a protective barrier against pathogen colonization. Microbes can engage in competitive interactions to eliminate pathogens by controlling access to space and nutrients through surface occupation and by secreting various metabolites. For example, the skin commensal *Staphylococcus epidermidis* engages in competitive interactions by secreting epidermin, a lantibiotic that acts against a broad spectrum of Gram-positive bacteria [72]. Microbes secrete volatile metabolites that can also function as signaling molecules, both within microbial communities, and between microbes and other organisms [86]. For example, VOCs released by the skin commensal *Staphylococcus epidermidis* have been found to be attractive to female malaria mosquitoes [83]. The role of VOCs released by different skin commensal bacterial species in bacteria-bacteria interactions is less well understood. Consideration must be given to the potential network of VOC-mediated interspecies communication that can influence and modulate the local microbiota, and in turn, affect cutaneous health. In this regard, Lemfack and co-workers investigated the volatile profiles of the 2 dominant families of skin microbiota, namely Staphylococcaceae and Corynebacteriaceae [78]. VOCs were collected in cell cultures using an adsorbent trap with subsequent solvent extraction and analysis by GC-MS. Principle component analysis distinguished the 2 bacterial families on the basis of their VOC profiles. 11 compounds were identified from *Corynebacterium*, including 2-nonanone, 2-phenylethanol, 2-undecanone and 2-pentadecanone, and >50 VOCs were identified from *Staphylococcus* strains, wherein ketones were predominant. One strain of *Staphylococcus* (*S. schleiferi*) was found to produce unique volatile profiles, with 2 newly identified substances (3-(2-phenylethylamino)butan-2-one and (*E*)-3-(2-phenylethylimino)butan-2-one) released in large quantities by one isolate of *S. schleiferi* (DSMZ4807). Co-culture experiments revealed that VOCs produced by the DSMZ4807 isolate selectively inhibited the growth of Gram-positive bacteria and influenced specific phenotypes (prodigiosin synthesis, bioluminescence) in Gram-negative bacteria [78]. This research highlights the potential role mVOCs play in bacteria-bacteria interactions among skin commensals and suggests that mVOCs might contribute to shaping the composition and diversity of the skin microbiota.

In addition to having distinct physiological effects on cohabiting or competing microbes, mVOCs are also known to affect the physiology of their hosts [75]. For instance, a high proportion of *Propionibacterium* typically colonizes sebaceous regions of skin where it hydrolyses the triglycerides in sebum to release beneficial fatty acids on the skin. These fatty acids act on the skin by acidifying and emolliating the local environment [65],[87]. On the other hand, evidence for the potential role mVOCs might play in disease processes in humans is beginning to emerge. Many mVOCs are considered as virulence factors, and they can have a variety of different actions on mammalian cells, including inflammatory and cytotoxic effects [88]. For instance, the production of short chain fatty acids (SCFAs) by the human gut microbiota can influence the physiology of the colon [89], and contribute to the development and progression of diseases like ulcerative colitis [90] and asthma [91]. Numerous VOCs produced by the metabolism of skin microorganisms are bioactive compounds, including SCFAs and many volatile amines (especially tryptamine and histamine) that can act as inflammatory mediators [92]. Further research is needed to understand the impact mVOCs might have on skin physiology, as well as the potential network of VOC-mediated interspecies communication and its influence on the local microbiota and host physiology. Skin microbial communities are highly sensitive to their underlying environment and its changing conditions, and their rapid response to a perturbation provides a sensitive readout of the environment within which they reside [93]. It is expected that differences in microbial metabolic processes would be reflected in volatile profiles, thus investigation of skin microbial communities *via* mVOCs could enlighten our understanding of their role in cutaneous health and disease, and potentially enable the manipulation of human skin microbiota through development of targeted pro-biotics [93].

2.3.2 Microbial volatiles as diagnostic tools for non-invasive detection of cutaneous wound infections

Microbial volatiles are considered promising as non-invasive diagnostic tools for earlier detection and treatment of microbial infections [94]. For centuries, the characteristic odour produced by *Clostridium perfringens* has been used in recognizing gas gangrene, a severe skin and soft tissue infection [95]. Clinical diagnosis of infections *via* analysis of mVOCs is currently used to diagnose pulmonary [96] and *Helicobacter pylori* infections [97], and there is a growing body of literature proposing volatile markers associated with *Staphylococcus aureus* and other pathogens commonly associated with lung and skin infections [98],[99]. Early identification of mVOCs in wound infections could offer a non-invasive, painless and reproducible diagnostic approach that could enable earlier detection of infection before the onset of malodour, chronicity or necrosis. Current wound infection diagnosis involves surface swabbing, wound exudate culture, wound tissue biopsy and the clinician's judgement of the signs of infection (erythema, oedema, warmth and purulence). Additional criteria used to judge signs of infection include discoloration, granulation and wound malodour (*i.e.* mVOCs). Current diagnostic approaches raise concerns over trauma, reproducibility and subjective observations [100],[101] and the signs of infection may not be apparent until an infection is well-established. This has motivated a growing research effort around the analysis of mVOCs for non-invasive diagnosis and management of cutaneous wound infections.

Complementary studies on skin VOCs from different sources (tissue and culture studies as well as *in vivo* research) will assist in validating the origin of candidate volatile compounds and their potential association with cutaneous health or disease. Recent evidence for the emission of unique VOC profiles during bacterial biofilm formation in human *ex vivo* cutaneous wound models has emerged (Table 1) [102]. Headspace samples were collected from biofilms of methicillin sensitive *Staphylococcus aureus* (MSSA), *Pseudomonas aeruginosa* (PA) and *Streptococcus pyogenes* (SP), grown in *in vitro* and *ex vivo* cutaneous wound organ culture models. The presence and abundance of specific VOCs was influenced by bacterial species, biofilm growth phase and wound model. For example, pentanal and 3-methylbutanal were uniquely identified in the early biofilm phases of MSSA in all wound models. One alkene (1-undecene) was present in both PA and SP biofilms in all wound models at all time points, while several other VOCs were uniquely identified in SP and PA biofilms respectively (Table 1). There were 2 VOCs (2-methyl-1-propanol and 3-methyl-1-butanol) common to all 3 bacteria, but differences in their relative abundance permitted differentiation between bacterial species [102]. These early results highlight the potential utility of VOCs in assessing biofilm development.

A number of non-invasive approaches for *in vivo* sampling and analysis of VOC profiles in the headspace of cutaneous wounds have emerged. Thomas *et al.* reported a non-invasive method for *in vivo* VOC sampling in chronic leg wounds [20]. Polydimethylsiloxane membranes were used to collect samples from 5 patients (from the wound, a control region of healthy skin, and a boundary region between the two) and analysed by gas chromatography ion trap mass spectrometry. They reported VOC profile differentiation between the control and lesion areas and between the control and boundary areas ($p<0.05$) as outlined in Table 1, but not between the lesion and boundary areas. Sensor arrays have developed as a portable approach to wound monitoring, and can differentiate between uninfected and infected venous leg wounds based on isolated beta-hemolytic *streptococci* [103]. An e-nose system comprising of metal oxide sensors with a SPME desorption unit for sample introduction was reported for monitoring burn patients *via* VOCs. The system distinguished between infected and uninfected burns based on samples obtained by SPME pre-concentration of VOCs from swabs and wound dressing material [104]. A commercial e-nose (Cyranose320) employing various classifiers (including linear discriminant analysis) demonstrated the capability to differentiate between single and poly-microbial species associated with diabetic foot infection under different agar culture conditions *in vitro* [105]. Further development of *in vivo* sampling approaches and identification of candidate biomarkers coupled with continued optimization and miniaturization of instrumentation, in addition to development of data analytics tools [106], may open up the possibility of point-of-care wound monitoring. It will be important to overcome the issue of the complexity of VOC profiles, related in part to the diversity of microbes, identifying commensal microbes and wounds that are unrelated to the pathogenic microbial communities present, in addition to the diversity within the pathogenic microbial communities. These challenges enforce the need for translational research in furthering the volatilomics field. Microbiome diagnostics *via* mVOCs may find application in personalized medicine for non-invasive diagnosis and management of cutaneous infection and dermatological disorders.

## 3. Conclusion

Studying the biochemical processes in the skin involving VOCs is a promising frontier for non-invasive research on skin physiology. It is a versatile approach enabling complementary investigations in tissue, cell culture and human participants. Successful deployment of sensor technologies for detection and classification of skin volatiles could be highly beneficial for human healthcare [107] in providing more objective and less-invasive examination methods to support clinicians in the decision making process, and enabling personalized management of cutaneous health and disease in the future. There is a wealth of opportunity for more research into the rich biochemistry of skin volatile secretions, particularly in the presence of specific disease states, however, there are still a number of analytical challenges to be overcome in the determination of skin volatiles in order to enable better comparability of results between laboratories and to inform the development of skin VOC-based diagnostics and sensor technologies. The chemical profile of skin volatiles varies depending on the sampling method used due to selective adsorption of analytes on different sampling materials. Furthermore, identification of volatile metabolites from skin is often solely reliant on comparing the mass spectrum of an unknown to the reference spectra in a commercial MS library. A move towards standardized methodology for sample collection and analysis would improve comparability between results generated in different laboratories. Investigation and optimisation of SPME adsorption, storage and desorption parameters will ensure a sensitive and reproducible method for skin VOC pre-concentration. Careful selection of experimental materials to reduce sample contamination (*e.g.* glass or inert polymers such as Teflon) should be considered. Isolation of the HS atmosphere (*e.g.* using a wearable sampler *in vivo* [23] or headspace vials *in vitro* [75]) could further reduce the incidence of environmental contaminants. Careful elucidation of endogenous volatile biomarkers with consideration of cosmetic and environmental influences is essential. GC-MS compound identification should be supported by independent data through retention index matching or reference standards [108] to reduce mistakes in VOC identification and reporting. There will be a need for VOC libraries specific to skin to permit rapid identification of compounds in a patient's sample [20]. This will likely be supported by investigations into the production of volatile metabolites at the cellular [49] and microbial [99] levels. Meeting these needs will support development of tailored recognition elements within sensor arrays for maximum selectivity and sensitivity based on pre-evaluated volatile profiles. Rapid advancements in sensing platforms are paving the way for a new class of wearable technology [109] with applications in cutaneous health emerging including wearables for wound monitoring [110], and epidermal sensors for skin hydration [111] and detection of sub-cutaneous inflammation [112]. This new class of wearable technology is already showing promise for targeted VOC sensing on skin [113] and has significance for a variety of application areas including clinical diagnostics, management of therapies, and fundamental cell biology and skin physiology research.

**Acknowledgements**

E. Duffy was supported by funding from the European Union's Horizon 2020 Research and Innovation programme under the Marie Skłodowska-Curie grant agreement number 796289. A. Morrin was supported by funding from Science Foundation Ireland under the Career Development Award grant number 13/CDA/2155.

**Table 1.** Summary of emerging volatile markers associated with cutaneous disease

| Disease/ disorder | n[1] | Sampling Method | Analytical method | Comments | Volatile compounds[2] | p- value |
|---|---|---|---|---|---|---|
| Compressed tissue (pressure ulcers) [22] | 18 | Cotton patch, HS-SPME; 15 hr extraction | GC-MS, E-nose | No definitive markers identified, rather different patterns of emission in healthy vs compressed tissue | 1,2,3-Propanetriol triacetate^ Tritetracontane^ | <0.1 |
| Acute barrier disruption (atopic dermatitis) [23] | 7 | HS-SPME of skin; 15 min extraction | GC-MS | Differential volatile emissions before/after impairment. Compound identities confirmed through RI matching | **Glycine**^ **Squalane*** **Octanal*** **Squalene*** **Nonanal*** | 0.0002 0.002 0.05 0.05 0.08 |
| Melanoma [21] | 2-40 | Gauze pad, HS-SPME; 15 hr extraction | GC-MS, E-nose | Subtle differences in GC-MS data for 2 participants, but compound identities not reported. E-nose melanoma prediction correct in 70% of cases. | Propanal | |
| Melanoma [60] | 8 | HS-SPME of tissue biopsies; 1 hr extraction | GC-MS | Differential expression of volatiles observed in melanoma vs matching skin control. 3 candidate biomarkers in frozen melanoma samples proposed. | 4-methyl decane^ Undecane^ Dodecane^ | <0.0001 <0.0001 0.0006 |
| Melanoma [62] | 10 | HS-SPME of tissue biopsies; 1 hr extraction | GC-MS | 6 significant features were identified for melanoma vs matching skin control. Confirmed through RI matching & standards. | **Dodecanoic acid**^ **Hexadecanoic acid**^ **Tetradecanoic acid**^ **1-Eicosene**^ **2-Ethylhexyl-4-methoxycinnamate**^ **Toluene*** | 0.05 |
| Melanoma cells [52] | | HS-SPME of cell culture, media & supernatant; 30 min | GC-MS | Quantitative comparison of 31 compounds showed significant differences between cell lines (melanoma & normal melanocytes). | **Dimethyl disulfide**[§] **Dimethyl trisulfide**[§] | <0.006 |

| | | | | | |
|---|---|---|---|---|---|
| Chronic arterial leg ulcers [20] | 5 | Polydimethylsiloxane membrane; 30 min extraction | Thermal desorption GC-MS | Differentiation between wound & lesion areas | 1-(1-Methyethoxy)-2-propanol<br>Dimethyl disulfide<br>3-Carene<br>3,5-Bis(1,1-dimethylethyl) phenol<br>**Dimethylsulfone**[^] | <0.05 |
| Biofilms in *ex vivo* wound models [102] | | 1 mL headspace samples collected in syringe | GC-MS | Volatiles specific to bacterial species, as well as biofilm and model-specific volatiles were observed | Pentanal[m]<br>3-Methylbutanal[m]<br>Hydrogen cyanide[p]<br>5-Methyl-2-hexanamine[p]<br>5-Methyl-2-heptanamine[p]<br>2-Nonanone[p]<br>2-Undecanone[p]<br>Ethanol[s]<br>2-Butanol[s] | <0.05 |

[1] Number of participants for *in vivo* studies; [2] Compounds in bold had their identification supported by retention index matching or analysis of reference standards. Those compounds deemed to be from exogenous sources [52] were excluded; [^] Up-regulation of a compound in disease state; [*] Down-regulation of a compound in disease state. [$] Compound only present in disease state; [m] compound unique to methicillin sensitive *Staphylococcus aureus*; [p] Compound unique to *Pseudomonas aeruginosa*; [s] Compound unique to *Streptococcus pyogenes*

**References**

[1] S. Paliwal, B.H. Hwang, K.Y. Tsai, S. Mitragotri, Diagnostic opportunities based on skin biomarkers, Eur. J. Pharm. Sci. 50 (2013) 546–556. doi:10.1016/j.ejps.2012.10.009.

[2] D. Rigopoulos, G. Larios, A. Katsambas, Skin signs of systemic diseases, Clin. Dermatol. 29 (2011) 531–540. doi:10.1016/j.clindermatol.2010.09.021.

[3] M. Shirasu, K. Touhara, The scent of disease: volatile organic compounds of the human body related to disease and disorder, J. Biochem. (Tokyo). 150 (2011) 257–266. doi:10.1093/jb/mvr090.

[4] S. Kippenberger, J. Havlíček, A. Bernd, D. Thaçi, R. Kaufmann, M. Meissner, "Nosing Around" the human skin: what information is concealed in skin odour?, Exp. Dermatol. 21 (2012) 655–659. doi:10.1111/j.1600-0625.2012.01545.x.

[5] B. de Lacy Costello, A. Amann, H. Al-Kateb, C. Flynn, W. Filipiak, T. Khalid, D. Osborne, N.M. Ratcliffe, A review of the volatiles from the healthy human body, J. Breath Res. 8 (2014) 014001. doi:10.1088/1752-7155/8/1/014001.

[6] E. Duffy, G. Albero, A. Morrin, Headspace Solid-Phase Microextraction Gas Chromatography-Mass Spectrometry Analysis of Scent Profiles from Human Skin, Cosmetics. 5 (2018) 62. doi:10.3390/cosmetics5040062.

[7] A. Caroprese, S. Gabbanini, C. Beltramini, E. Lucchi, L. Valgimigli, HS-SPME-GC-MS analysis of body odor to test the efficacy of foot deodorant formulations, Skin Res. Technol. 15 (2009) 503–510. doi:10.1111/j.1600-0846.2009.00399.x.

[8] Y.Y. Broza, P. Mochalski, V. Ruzsanyi, A. Amann, H. Haick, Hybrid volatolomics and disease detection, Angew. Chem. 54 (2015) 11036–11048. doi:10.1002/anie.201500153.

[9] P. Prada, K. Furton, Human Scent Detection: A Review of its Developments and Forensic Applications, Rev. Cienc. Forenses. 1 (2008) 81–87.

[10] P. Mochalski, V. Ruzsanyi, H. Wiesenhofer, C.A. Mayhew, Instrumental sensing of trace volatiles-a new promising tool for detecting the presence of entrapped or hidden people, J. Breath Res. 12 (2018) 027107. doi:10.1088/1752-7163/aa9769.

[11] H.J. Martin, M.A. Turner, S. Bandelow, L. Edwards, S. Riazanskaia, C.L.P. Thomas, Volatile organic compound markers of psychological stress in skin: a pilot study, J. Breath Res. 10 (2016) 046012. doi:10.1088/1752-7155/10/4/046012.

[12] U.R. Bernier, D.L. Kline, D.R. Barnard, C.E. Schreck, R.A. Yost, Analysis of human skin emanations by gas chromatography/mass spectrometry. 2. Identification of volatile compounds that are candidate attractants for the yellow fever mosquito (Aedes aegypti), Anal. Chem. 72 (2000) 747–756.

[13] C. Turner, B. Parekh, C. Walton, P. Spanel, D. Smith, M. Evans, An exploratory comparative study of volatile compounds in exhaled breath and emitted by skin using selected ion flow tube mass spectrometry, Rapid Commun. Mass Spectrom. 22 (2008) 526–532. doi:10.1002/rcm.3402.

[14] R. Mebazaa, B. Rega, V. Camel, Analysis of human male armpit sweat after fenugreek ingestion: Characterisation of odour active compounds by gas chromatography coupled to mass spectrometry and olfactometry, Food Chem. 128 (2011) 227–235. doi:10.1016/j.foodchem.2011.02.063.

[15] J. Havlicek, T. Saxton, New Research on Food Habits, in: Eff. Diet Hum. Bodily Odors, Nova Science Publishers, 2009: pp. 35–44.

[16] M. Gallagher, C.J. Wysocki, J.J. Leyden, A.I. Spielman, X. Sun, G. Preti, Analyses of volatile organic compounds from human skin, Br. J. Dermatol. 159 (2008) 780–791. doi:10.1111/j.1365-2133.2008.08748.x.

[17] C. Turner, Potential of breath and skin analysis for monitoring blood glucose concentration in diabetes, Expert Rev. Mol. Diagn. 11 (2011) 497–503. doi:10.1586/erm.11.31.

[18] K. Nose, T. Mizuno, N. Yamane, T. Kondo, H. Ohtani, S. Araki, T. Tsuda, Identification of ammonia in gas emanated from human skin and its correlation with that in blood, Anal. Sci. Int. J. Jpn. Soc. Anal. Chem. 21 (2005) 1471–1474.

[19] C.M. De Moraes, C. Wanjiku, N.M. Stanczyk, H. Pulido, J.W. Sims, H.S. Betz, A.F. Read, B. Torto, M.C. Mescher, Volatile biomarkers of symptomatic and asymptomatic malaria infection in humans, Proc. Natl. Acad. Sci. U. S. A. 115 (2018) 5780–5785. doi:10.1073/pnas.1801512115.

[20] A.N. Thomas, S. Riazanskaia, W. Cheung, Y. Xu, R. Goodacre, C.L.P. Thomas, M.S. Baguneid, A. Bayat, Novel noninvasive identification of biomarkers by analytical profiling of chronic wounds using volatile organic compounds, Wound Repair Regen. 18 (2010) 391–400. doi:10.1111/j.1524-475X.2010.00592.x.

[21] A. D'Amico, R. Bono, G. Pennazza, M. Santonico, G. Mantini, M. Bernabei, M. Zarlenga, C. Roscioni, E. Martinelli, R. Paolesse, C. Di Natale, Identification of melanoma with a gas sensor array, Skin Res. Technol. Off. J. Int. Soc. Bioeng. Skin ISBS Int. Soc. Digit. Imaging Skin ISDIS Int. Soc. Skin Imaging ISSI. 14 (2008) 226–236. doi:10.1111/j.1600-0846.2007.00284.x.

[22] F. Dini, R. Capuano, T. Strand, A.-C. Ek, M. Lindgren, R. Paolesse, C. Di Natale, I. Lundström, Volatile Emissions from Compressed Tissue, PLoS ONE. 8 (2013). doi:10.1371/journal.pone.0069271.

[23] E. Duffy, M.R. Jacobs, B. Kirby, A. Morrin, Probing skin physiology through the volatile footprint: Discriminating volatile emissions before and after acute barrier disruption, Exp. Dermatol. 26 (2017) 919–925. doi:10.1111/exd.13344.

[24] H. Kataoka, K. Saito, H. Kato, K. Masuda, Noninvasive analysis of volatile biomarkers in human emanations for health and early disease diagnosis, Bioanalysis. 5 (2013) 1443–1459. doi:10.4155/bio.13.85.

[25] M.M.L. Steeghs, B.W.M. Moeskops, K. van Swam, S.M. Cristescu, P.T.J. Scheepers, F.J.M. Harren, On-line monitoring of UV-induced lipid peroxidation products from human skin in vivo using proton-transfer reaction mass spectrometry, Int. J. Mass Spectrom. 253 (2006) 58–64. doi:10.1016/j.ijms.2006.02.015.

[26] S. Giannoukos, B. Brkić, S. Taylor, N. France, Monitoring of human chemical signatures using membrane inlet mass spectrometry, Anal. Chem. 86 (2014) 1106–1114. doi:10.1021/ac403621c.

[27] P. Martínez-Lozano, Mass spectrometric study of cutaneous volatiles by secondary electrospray ionization, Int. J. Mass Spectrom. 282 (2009) 128–132. doi:10.1016/j.ijms.2009.02.017.

[28] V. Ruzsanyi, P. Mochalski, A. Schmid, H. Wiesenhofer, M. Klieber, H. Hinterhuber, A. Amann, Ion mobility spectrometry for detection of skin volatiles, J. Chromatogr. B Analyt. Technol. Biomed. Life. Sci. 911 (2012) 84–92. doi:10.1016/j.jchromb.2012.10.028.

[29] P. Mochalski, H. Wiesenhofer, M. Allers, S. Zimmermann, A.T. Güntner, N.J. Pineau, W. Lederer, A. Agapiou, C.A. Mayhew, V. Ruzsanyi, Monitoring of selected skin- and breath-borne volatile organic compounds emitted from the human body using gas chromatography ion mobility spectrometry (GC-IMS), J. Chromatogr. B. 1076 (2018) 29–34. doi:10.1016/j.jchromb.2018.01.013.

[30] C. Wongchoosuk, M. Lutz, T. Kerdcharoen, Detection and Classification of Human Body Odor Using an Electronic Nose, Sensors. 9 (2009) 7234–7249. doi:10.3390/s90907234.

[31] W.J. Harper, The strengths and weaknesses of the electronic nose, Adv. Exp. Med. Biol. 488 (2001) 59–71.

[32] L. Dormont, J.-M. Bessière, D. McKey, A. Cohuet, New methods for field collection of human skin volatiles and perspectives for their application in the chemical ecology of human-pathogen-vector interactions, J. Exp. Biol. 216 (2013) 2783–2788. doi:10.1242/jeb.085936.

[33] A.M. Curran, S.I. Rabin, P.A. Prada, K.G. Furton, Comparison of the volatile organic compounds present in human odor using SPME-GC/MS, J. Chem. Ecol. 31 (2005) 1607–1619.

[34] P. Mochalski, J. King, K. Unterkofler, H. Hinterhuber, A. Amann, Emission rates of selected volatile organic compounds from skin of healthy volunteers, J. Chromatogr. B. 959 (2014) 62–70. doi:10.1016/j.jchromb.2014.04.006.

[35] L. Dormont, J.-M. Bessière, A. Cohuet, Human skin volatiles: a review, J. Chem. Ecol. 39 (2013) 569–578. doi:10.1007/s10886-013-0286-z.

[36]   S. Riazanskaia, G. Blackburn, M. Harker, D. Taylor, C.L.P. Thomas, The analytical utility of thermally desorbed polydimethylsilicone membranes for in-vivo sampling of volatile organic compounds in and on human skin, The Analyst. 133 (2008) 1020–1027. doi:10.1039/b802515k.

[37]   R. Jiang, E. Cudjoe, B. Bojko, T. Abaffy, J. Pawliszyn, A non-invasive method for in vivo skin volatile compounds sampling, Anal. Chim. Acta. 804 (2013) 111–119. doi:10.1016/j.aca.2013.09.056.

[38]   A.P. Roodt, Y. Naudé, A. Stoltz, E. Rohwer, Human skin volatiles: Passive sampling and GC × GC-ToFMS analysis as a tool to investigate the skin microbiome and interactions with anthropophilic mosquito disease vectors, J. Chromatogr. B Analyt. Technol. Biomed. Life. Sci. 1097–1098 (2018) 83–93. doi:10.1016/j.jchromb.2018.09.002.

[39]   A.M. Curran, P.A. Prada, K.G. Furton, The differentiation of the volatile organic signatures of individuals through SPME-GC/MS of characteristic human scent compounds, J. Forensic Sci. 55 (2010) 50–57. doi:10.1111/j.1556-4029.2009.01236.x.

[40]   J.G. Logan, M.A. Birkett, S.J. Clark, S. Powers, N.J. Seal, L.J. Wadhams, A.J. Mordue Luntz, J.A. Pickett, Identification of human-derived volatile chemicals that interfere with attraction of Aedes aegypti mosquitoes, J. Chem. Ecol. 34 (2008) 308–322. doi:10.1007/s10886-008-9436-0.

[41]   K. Saga, Structure and function of human sweat glands studied with histochemistry and cytochemistry, Prog. Histochem. Cytochem. 37 (2002) 323–386.

[42]   A. Watabe, T. Sugawara, K. Kikuchi, K. Yamasaki, S. Sakai, S. Aiba, Sweat constitutes several natural moisturizing factors, lactate, urea, sodium, and potassium, J. Dermatol. Sci. 72 (2013) 177–182. doi:10.1016/j.jdermsci.2013.06.005.

[43]   C.-Y. Cui, D. Schlessinger, Eccrine sweat gland development and sweat secretion, Exp. Dermatol. 24 (2015) 644–650. doi:10.1111/exd.12773.

[44]   J. Marshall, K.T. Holland, E.M. Gribbon, A comparative study of the cutaneous microflora of normal feet with low and high levels of odour, J. Appl. Bacteriol. 65 (1988) 61–68.

[45]   Z. Syed, W.S. Leal, Acute olfactory response of Culex mosquitoes to a human- and bird-derived attractant, Proc. Natl. Acad. Sci. U. S. A. 106 (2009) 18803–18808. doi:10.1073/pnas.0906932106.

[46]   M. Kusano, E. Mendez, K.G. Furton, Comparison of the volatile organic compounds from different biological specimens for profiling potential, J. Forensic Sci. 58 (2013) 29–39. doi:10.1111/j.1556-4029.2012.02215.x.

[47]   S. Haze, Y. Gozu, S. Nakamura, Y. Kohno, K. Sawano, H. Ohta, K. Yamazaki, 2-Nonenal newly found in human body odor tends to increase with aging, J. Invest. Dermatol. 116 (2001) 520–524. doi:10.1046/j.0022-202x.2001.01287.x.

[48]   D.J. Penn, E. Oberzaucher, K. Grammer, G. Fischer, H.A. Soini, D. Wiesler, M.V. Novotny, S.J. Dixon, Y. Xu, R.G. Brereton, Individual and gender fingerprints in human body odour, J. R. Soc. Interface. 4 (2007) 331–340. doi:10.1098/rsif.2006.0182.

[49]   W. Filipiak, P. Mochalski, A. Filipiak, C. Ager, R. Cumeras, C.E. Davis, A. Agapiou, K. Unterkofler, J. Troppmair, A Compendium of Volatile Organic Compounds (VOCs) Released By Human Cell Lines, Curr. Med. Chem. 23 (2016) 2112–2131.

[50]   E. Duffy, K.D. Guzman, R. Wallace, R. Murphy, A. Morrin, E. Duffy, K.D. Guzman, R. Wallace, R. Murphy, A. Morrin, Non-Invasive Assessment of Skin Barrier Properties: Investigating Emerging Tools for In Vitro and In Vivo Applications, Cosmetics. 4 (2017) 44. doi:10.3390/cosmetics4040044.

[51]   C.A. Acevedo, E.Y. Sánchez, J.G. Reyes, M.E. Young, Volatile organic compounds produced by human skin cells, Biol. Res. 40 (2007) 347–355. doi:/S0716-97602007000400009.

[52]   J. Kwak, M. Gallagher, M.H. Ozdener, C.J. Wysocki, B.R. Goldsmith, A. Isamah, A. Faranda, S.S. Fakharzadeh, M. Herlyn, A.T.C. Johnson, G. Preti, Volatile biomarkers from human melanoma cells, J. Chromatogr. B. 931 (2013) 90–96. doi:10.1016/j.jchromb.2013.05.007.

[53]   V.K. Wong, N.A. Stotts, H.W. Hopf, E.S. Froelicher, G.A. Dowling, How heel oxygenation changes under pressure, Wound Repair Regen. 15 (2007) 786–794. doi:10.1111/j.1524-475X.2007.00309.x.

[54] M. Denda, New strategies to improve skin barrier homeostasis, Adv. Drug Deliv. Rev. 54 Suppl 1 (2002) S123-130.

[55] H. Williams, A. Pembroke, Sniffer dogs in the melanoma clinic?, Lancet Lond. Engl. 1 (1989) 734.

[56] J. Church, H. Williams, Another sniffer dog for the clinic?, Lancet Lond. Engl. 358 (2001) 930. doi:10.1016/S0140-6736(01)06065-2.

[57] C.M. Willis, L.E. Britton, M.A. Swindells, E.M. Jones, A.E. Kemp, N.L. Muirhead, A. Gul, R.N. Matin, L. Knutsson, M. Ali, Invasive melanoma in vivo can be distinguished from basal cell carcinoma, benign naevi and healthy skin by canine olfaction: a proof-of-principle study of differential volatile organic compound emission, Br. J. Dermatol. 175 (2016) 1020–1029. doi:10.1111/bjd.14887.

[58] J. Dinnes, R.N. Matin, J.F. Moreau, L. Patel, S.A. Chan, N. Chuchu, S.E. Bayliss, M. Grainge, Y. Takwoingi, C. Davenport, F.M. Walter, C. Fleming, J. Schofield, N. Shroff, K. Godfrey, C. O'Sullivan, J.J. Deeks, H.C. Williams, Tests to assist in the diagnosis of cutaneous melanoma in adults: a generic protocol, Cochrane Database Syst. Rev. (2015). doi:10.1002/14651858.CD011902.

[59] D. Pickel, G.P. Manucy, D.B. Walker, S.B. Hall, J.C. Walker, Evidence for canine olfactory detection of melanoma, Appl. Anim. Behav. Sci. 89 (2004) 107–116. doi:10.1016/j.applanim.2004.04.008.

[60] T. Abaffy, R. Duncan, D.D. Riemer, O. Tietje, G. Elgart, C. Milikowski, R.A. DeFazio, Differential Volatile Signatures from Skin, Naevi and Melanoma: A Novel Approach to Detect a Pathological Process, PLoS ONE. 5 (2010). doi:10.1371/journal.pone.0013813.

[61] C. Fink, H.A. Haenssle, Non-invasive tools for the diagnosis of cutaneous melanoma, Skin Res. Technol. 23 (2017) 261–271. doi:10.1111/srt.12350.

[62] T. Abaffy, M.G. Möller, D.D. Riemer, C. Milikowski, R.A. DeFazio, Comparative analysis of volatile metabolomics signals from melanoma and benign skin: a pilot study, Metabolomics. 9 (2013) 998–1008. doi:10.1007/s11306-013-0523-z.

[63] A. Bartolazzi, M. Santonico, G. Pennazza, E. Martinelli, R. Paolesse, A. D'Amico, C. Di Natale, A sensor array and GC study about VOCs and cancer cells, Sens. Actuators B Chem. 146 (2010) 483–488. doi:10.1016/j.snb.2009.11.046.

[64] Z. Wang, M. Sun, C. Wang, Detection of Melanoma Cancer Biomarker Dimethyl Disulfide Using Cavity Ringdown Spectroscopy at 266 nm, Appl. Spectrosc. 70 (2016) 1080–1085. doi:10.1177/0003702816641575.

[65] E.A. Grice, H.H. Kong, S. Conlan, C.B. Deming, J. Davis, A.C. Young, G.G. Bouffard, R.W. Blakesley, P.R. Murray, E.D. Green, M.L. Turner, J.A. Segre, Topographical and Temporal Diversity of the Human Skin Microbiome, Science. 324 (2009) 1190–1192. doi:10.1126/science.1171700.

[66] A.L. Cogen, V. Nizet, R.L. Gallo, Skin microbiota: a source of disease or defence?, Br. J. Dermatol. 158 (2008) 442–455. doi:10.1111/j.1365-2133.2008.08437.x.

[67] B.S. Baker, The role of microorganisms in atopic dermatitis, Clin. Exp. Immunol. 144 (2006) 1–9. doi:10.1111/j.1365-2249.2005.02980.x.

[68] K.T. Holland, E. Ingham, W.J. Cunliffe, A review, the microbiology of acne, J. Appl. Bacteriol. 51 (1981) 195–215.

[69] R.L. Gallo, T. Nakatsuji, Microbial symbiosis with the innate immune defense system of the skin, J. Invest. Dermatol. 131 (2011) 1974–1980. doi:10.1038/jid.2011.182.

[70] H.H. Kong, J. Oh, C. Deming, S. Conlan, E.A. Grice, M.A. Beatson, E. Nomicos, E.C. Polley, H.D. Komarow, NISC Comparative Sequence Program, P.R. Murray, M.L. Turner, J.A. Segre, Temporal shifts in the skin microbiome associated with disease flares and treatment in children with atopic dermatitis, Genome Res. 22 (2012) 850–859. doi:10.1101/gr.131029.111.

[71] S. Naik, N. Bouladoux, C. Wilhelm, M.J. Molloy, R. Salcedo, W. Kastenmuller, C. Deming, M. Quinones, L. Koo, S. Conlan, S. Spencer, J.A. Hall, A. Dzutsev, H. Kong, D.J. Campbell, G. Trinchieri, J.A. Segre, Y. Belkaid, Compartmentalized control of skin immunity by resident commensals, Science. 337 (2012) 1115–1119. doi:10.1126/science.1225152.

[72] F. Götz, S. Perconti, P. Popella, R. Werner, M. Schlag, Epidermin and gallidermin: Staphylococcal lantibiotics, Int. J. Med. Microbiol. IJMM. 304 (2014) 63–71. doi:10.1016/j.ijmm.2013.08.012.

[73] S. Schulz, J.S. Dickschat, Bacterial volatiles: the smell of small organisms, Nat. Prod. Rep. 24 (2007) 814–842. doi:10.1039/b507392h.

[74] C. Heddergott, A.M. Calvo, J.P. Latgé, The volatome of Aspergillus fumigatus, Eukaryot. Cell. 13 (2014) 1014–1025. doi:10.1128/EC.00074-14.

[75] C.M. Timm, E.P. Lloyd, A. Egan, R. Mariner, D. Karig, Direct Growth of Bacteria in Headspace Vials Allows for Screening of Volatiles by Gas Chromatography Mass Spectrometry, Front. Microbiol. 9 (2018) 491. doi:10.3389/fmicb.2018.00491.

[76] A.G. James, C.J. Austin, D.S. Cox, D. Taylor, R. Calvert, Microbiological and biochemical origins of human axillary odour, FEMS Microbiol. Ecol. 83 (2013) 527–540. doi:10.1111/1574-6941.12054.

[77] K. Ara, M. Hama, S. Akiba, K. Koike, K. Okisaka, T. Hagura, T. Kamiya, F. Tomita, Foot odor due to microbial metabolism and its control, Can. J. Microbiol. 52 (2006) 357–364. doi:10.1139/w05-130.

[78] M.C. Lemfack, S.R. Ravella, N. Lorenz, M. Kai, K. Jung, S. Schulz, B. Piechulla, Novel volatiles of skin-borne bacteria inhibit the growth of Gram-positive bacteria and affect quorum-sensing controlled phenotypes of Gram-negative bacteria, Syst. Appl. Microbiol. 39 (2016) 503–515. doi:10.1016/j.syapm.2016.08.008.

[79] A.G. James, D. Hyliands, H. Johnston, Generation of volatile fatty acids by axillary bacteria, Int. J. Cosmet. Sci. 26 (2004) 149–156. doi:10.1111/j.1467-2494.2004.00214.x.

[80] D. Stevens, R. Cornmell, D. Taylor, S.G. Grimshaw, S. Riazanskaia, D.S. Arnold, S.J. Fernstad, A.M. Smith, L.M. Heaney, J.C. Reynolds, C.L.P. Thomas, M. Harker, Spatial variations in the microbial community structure and diversity of the human foot is associated with the production of odorous volatiles, FEMS Microbiol. Ecol. 91 (2015) 1–11. doi:10.1093/femsec/fiu018.

[81] N.O. Verhulst, H. Beijleveld, B.G. Knols, W. Takken, G. Schraa, H.J. Bouwmeester, R.C. Smallegange, Cultured skin microbiota attracts malaria mosquitoes, Malar. J. 8 (2009) 302. doi:10.1186/1475-2875-8-302.

[82] N.O. Verhulst, R. Andriessen, U. Groenhagen, G. Bukovinszkiné Kiss, S. Schulz, W. Takken, J.J.A. van Loon, G. Schraa, R.C. Smallegange, Differential attraction of malaria mosquitoes to volatile blends produced by human skin bacteria, PloS One. 5 (2010) e15829. doi:10.1371/journal.pone.0015829.

[83] N.O. Verhulst, Y.T. Qiu, H. Beijleveld, C. Maliepaard, D. Knights, S. Schulz, D. Berg-Lyons, C.L. Lauber, W. Verduijn, G.W. Haasnoot, R. Mumm, H.J. Bouwmeester, F.H.J. Claas, M. Dicke, J.J.A. van Loon, W. Takken, R. Knight, R.C. Smallegange, Composition of human skin microbiota affects attractiveness to malaria mosquitoes, PloS One. 6 (2011) e28991. doi:10.1371/journal.pone.0028991.

[84] M. Ashrafi, M. Bates, M. Baguneid, T. Alonso-Rasgado, R. Rautemaa-Richardson, A. Bayat, Volatile organic compound detection as a potential means of diagnosing cutaneous wound infections, Wound Repair Regen. 25 (2017) 574–590. doi:10.1111/wrr.12563.

[85] E.E. Geldreich, B.A. Kenner, P.W. Kabler, Occurrence of Coliforms, Fecal Coliforms, and Streptococci on Vegetation and Insects, Appl. Microbiol. 12 (1964) 63–69.

[86] B. Audrain, M.A. Farag, C.-M. Ryu, J.-M. Ghigo, Role of bacterial volatile compounds in bacterial biology, FEMS Microbiol. Rev. 39 (2015) 222–233. doi:10.1093/femsre/fuu013.

[87] R.R. Marples, D.T. Downing, A.M. Kligman, Control of free fatty acids in human surface lipids by Corynebacterium acnes, J. Invest. Dermatol. 56 (1971) 127–131.

[88] R.M.S. Thorn, J. Greenman, Microbial volatile compounds in health and disease conditions, J. Breath Res. 6 (2012) 024001. doi:10.1088/1752-7155/6/2/024001.

[89] D. Ríos-Covián, P. Ruas-Madiedo, A. Margolles, M. Gueimonde, C.G. de los Reyes-Gavilán, N. Salazar, Intestinal Short Chain Fatty Acids and their Link with Diet and Human Health, Front. Microbiol. 7 (2016). doi:10.3389/fmicb.2016.00185.

[90] K. Machiels, M. Joossens, J. Sabino, V. De Preter, I. Arijs, V. Eeckhaut, V. Ballet, K. Claes, F. Van Immerseel, K. Verbeke, M. Ferrante, J. Verhaegen, P. Rutgeerts, S. Vermeire, A decrease

of the butyrate-producing species Roseburia hominis and Faecalibacterium prausnitzii defines dysbiosis in patients with ulcerative colitis, Gut. 63 (2014) 1275–1283. doi:10.1136/gutjnl-2013-304833.

[91] M.-C. Arrieta, L.T. Stiemsma, P.A. Dimitriu, L. Thorson, S. Russell, S. Yurist-Doutsch, B. Kuzeljevic, M.J. Gold, H.M. Britton, D.L. Lefebvre, P. Subbarao, P. Mandhane, A. Becker, K.M. McNagny, M.R. Sears, T. Kollmann, CHILD Study Investigators, W.W. Mohn, S.E. Turvey, B.B. Finlay, Early infancy microbial and metabolic alterations affect risk of childhood asthma, Sci. Transl. Med. 7 (2015) 307ra152. doi:10.1126/scitranslmed.aab2271.

[92] R.P. Allaker, J. Greenman, R.H. Osborne, The production of inflammatory compounds by Propionibacterium acnes and other skin organisms, Br. J. Dermatol. 117 (1987) 175–183.

[93] E.A. Grice, The skin microbiome: potential for novel diagnostic and therapeutic approaches to cutaneous disease, Semin. Cutan. Med. Surg. 33 (2014) 98–103.

[94] M. Sohrabi, L. Zhang, K. Zhang, A. Ahmetagic, M.Q. Wei, Volatile Organic Compounds as Novel Markers for the Detection of Bacterial Infections, Clin. Microbiol. Open Access. 3 (2014). doi:10.4172/2327-5073.1000151.

[95] L.R. Bijland, M.K. Bomers, Y.M. Smulders, Smelling the diagnosis: a review on the use of scent in diagnosing disease, Neth. J. Med. 71 (2013) 300–307.

[96] J.E. Graham, Chapter Two - Bacterial Volatiles and Diagnosis of Respiratory Infections, in: S. Sariaslani, G.M. Gadd (Eds.), Adv. Appl. Microbiol., Academic Press, 2013: pp. 29–52. doi:10.1016/B978-0-12-407679-2.00002-8.

[97] K.M. Paschke, A. Mashir, R.A. Dweik, Clinical applications of breath testing, F1000 Med. Rep. 2 (2010). doi:10.3410/M2-56.

[98] W. Filipiak, A. Sponring, M.M. Baur, A. Filipiak, C. Ager, H. Wiesenhofer, M. Nagl, J. Troppmair, A. Amann, Molecular analysis of volatile metabolites released specifically by Staphylococcus aureus and Pseudomonas aeruginosa, BMC Microbiol. 12 (2012) 113. doi:10.1186/1471-2180-12-113.

[99] M.C. Lemfack, B.-O. Gohlke, S.M.T. Toguem, S. Preissner, B. Piechulla, R. Preissner, mVOC 2.0: a database of microbial volatiles, Nucleic Acids Res. 46 (2018) D1261–D1265. doi:10.1093/nar/gkx1016.

[100] S.E. Gardner, R.A. Frantz, C.L. Saltzman, S.L. Hillis, H. Park, M. Scherubel, Diagnostic validity of three swab techniques for identifying chronic wound infection, Wound Repair Regen. 14 (2006) 548–557. doi:10.1111/j.1743-6109.2006.00162.x.

[101] S.E. Gardner, R.A. Frantz, B.N. Doebbeling, The validity of the clinical signs and symptoms used to identify localized chronic wound infection, Wound Repair Regen. 9 (2001) 178–186.

[102] M. Ashrafi, L. Novak-Frazer, M. Bates, M. Baguneid, T. Alonso-Rasgado, G. Xia, R. Rautemaa-Richardson, A. Bayat, Validation of biofilm formation on human skin wound models and demonstration of clinically translatable bacteria-specific volatile signatures, Sci. Rep. 8 (2018) 9431. doi:10.1038/s41598-018-27504-z.

[103] A.D. Parry, P.R. Chadwick, D. Simon, B. Oppenheim, C.N. McCollum, Leg ulcer odour detection identifies beta-haemolytic streptococcal infection, J. Wound Care. 4 (1995) 404–406.

[104] H.-G. Byun, K. Persaud, A.M. Pisanelli, Wound-state monitoring for burn patients using E-Nose/SPME system, ETRI J. 32 (2010) 440–446. doi:10.4218/etrij.10.0109.0300.

[105] N. Yusuf, A. Zakaria, M.I. Omar, A.Y.M. Shakaff, M.J. Masnan, L.M. Kamarudin, N. Abdul Rahim, N.Z.I. Zakaria, A.A. Abdullah, A. Othman, M.S. Yasin, In-vitro diagnosis of single and poly microbial species targeted for diabetic foot infection using e-nose technology, BMC Bioinformatics. 16 (2015) 158. doi:10.1186/s12859-015-0601-5.

[106] S.I.C.J. Palma, A.P. Traguedo, A.R. Porteira, M.J. Frias, H. Gamboa, A.C.A. Roque, Machine learning for the meta-analyses of microbial pathogens' volatile signatures, Sci. Rep. 8 (2018) 3360. doi:10.1038/s41598-018-21544-1.

[107] R. Vishinkin, H. Haick, Nanoscale Sensor Technologies for Disease Detection via Volatolomics, Small Weinh. Bergstr. Ger. 11 (2015) 6142–6164. doi:10.1002/smll.201501904.

[108] L.W. Sumner, A. Amberg, D. Barrett, M.H. Beale, R. Beger, C.A. Daykin, T.W.-M. Fan, O. Fiehn, R. Goodacre, J.L. Griffin, T. Hankemeier, N. Hardy, J. Harnly, R. Higashi, J. Kopka, A.N. Lane, J.C. Lindon, P. Marriott, A.W. Nicholls, M.D. Reily, J.J. Thaden, M.R. Viant, Proposed minimum reporting standards for chemical analysis Chemical Analysis Working

Group (CAWG) Metabolomics Standards Initiative (MSI), Metabolomics Off. J. Metabolomic Soc. 3 (2007) 211–221. doi:10.1007/s11306-007-0082-2.

[109] H. Jin, Y.S. Abu-Raya, H. Haick, Advanced Materials for Health Monitoring with Skin-Based Wearable Devices, Adv. Healthc. Mater. 6 (2017). doi:10.1002/adhm.201700024.

[110] M.S. Brown, B. Ashley, A. Koh, Wearable Technology for Chronic Wound Monitoring: Current Dressings, Advancements, and Future Prospects, Front. Bioeng. Biotechnol. 6 (2018). doi:10.3389/fbioe.2018.00047.

[111] K. De Guzman, A. Morrin, Screen‐printed Tattoo Sensor towards the Non‐invasive Assessment of the Skin Barrier, Electroanalysis. 29 (2016). doi:10.1002/elan.201600572.

[112] S.R. Madhvapathy, Y. Ma, M. Patel, S. Krishnan, C. Wei, Y. Li, S. Xu, X. Feng, Y. Huang, J.A. Rogers, Epidermal Electronic Systems for Measuring the Thermal Properties of Human Skin at Depths of up to Several Millimeters, Adv. Funct. Mater. 28 (2018) 1802083. doi:10.1002/adfm.201802083.

[113] H. Jin, T.-P. Huynh, H. Haick, Self-Healable Sensors Based Nanoparticles for Detecting Physiological Markers via Skin and Breath: Toward Disease Prevention via Wearable Devices, Nano Lett. 16 (2016) 4194–4202. doi:10.1021/acs.nanolett.6b01066.